\documentclass[letterpaper]{article} 
\usepackage{aaai25}  
\usepackage{times}  
\usepackage{helvet}  
\usepackage{courier}  
\usepackage[hyphens]{url}  
\usepackage{graphicx} 
\usepackage{natbib}  
\usepackage{caption} 
\usepackage[dvipsnames]{xcolor}
\usepackage{algorithm}
\usepackage{algorithmic}
\usepackage{amsmath} 
\usepackage{booktabs}
\usepackage{tabularx,ragged2e}
\newcolumntype{L}{>{\arraybackslash}X} 
\usepackage{newfloat}
\usepackage{listings}
\DeclareCaptionStyle{ruled}{labelfont=normalfont,labelsep=colon,strut=off} 
\floatstyle{ruled}
\newfloat{listing}{tb}{lst}{}
\floatname{listing}{Listing}
\nocopyright 

\usepackage{color}
\usepackage{xcolor}
\usepackage{subcaption}
\usepackage{pifont}
\usepackage{tikz}
\usetikzlibrary{arrows.meta,chains,positioning,shapes.geometric}
\usepackage{marvosym}
\usepackage{fontawesome}
\usepackage[skins]{tcolorbox}
\DeclareUnicodeCharacter{0308}{o}

\title{No Such Thing as Free Brain Time: For a Pigouvian Tax on Attention Capture}

\author {
    Hamza Belgroun\textsuperscript{\rm 1},
    Franck Michel\textsuperscript{\rm 2},
    Fabien Gandon\textsuperscript{\rm 3}
}
\affiliations {
    \textsuperscript{\rm 1}Sciences Po, Inria, Université Côte d'Azur, CNRS\\
    \textsuperscript{\rm 2}University Côte d'Azur, CNRS, Inria\\
    \textsuperscript{\rm 3}Inria, University Côte d'Azur, CNRS\\
    hamza.belgroun31@gmail.com, franck.michel@inria.fr, fabien.gandon@inria.fr
}

\begin{document}

\maketitle

\begin{abstract}
In our age of digital platforms, human attention has become a scarce and highly valuable resource, rivalrous, tradable, and increasingly subject to market dynamics. This article explores the commodification of attention within the framework of the attention economy, arguing that attention should be understood as a common good threatened by over-exploitation. Drawing from philosophical, economic, and legal perspectives, we first conceptualize attention not only as an individual cognitive process but as a collective and infrastructural phenomenon susceptible to enclosure by digital intermediaries. 
We then identify and analyze negative externalities of the attention economy, particularly those stemming from excessive screen time: diminished individual agency, adverse health outcomes, and societal and political harms, including democratic erosion and inequality. These harms are largely unpriced by market actors and constitute a significant market failure. 
In response, among a spectrum of public policy tools ranging from informational campaigns to outright restrictions, we propose a Pigouvian tax on attention capture as a promising regulatory instrument to internalize the externalities and, in particular, the social cost of compulsive digital engagement. Such a tax would incentivize structural changes in platform design while preserving user autonomy. By reclaiming attention as a shared resource vital to human agency, health, and democracy, this article contributes a novel economic and policy lens to the debate on digital regulation. Ultimately, this article advocates for a paradigm shift: from treating attention as a private, monetizable asset to protecting it as a collective resource vital for humanity.
\end{abstract}


\section{The Nature of Attention}
\label{sec:nature_of_attention}

If you read these lines, we just won a fierce competition: we successfully attracted your attention against the many other agents attempting to capture it, and we were able to do so for free. Although we are used to being asked to “pay attention” when someone requests us to focus on something~\cite{legarrec2019price}, framing attention as a currency has never been more relevant than in the attention economy age~\cite{richards2023attention}. Produced by individuals as a limited resource, attention is then concentrated by an intermediary platform (e.g., a newspaper, social media), and subsequently sold to advertising companies, ready for consumption. Yet, the commodification of human attention is not a new phenomenon. It traces back to the nineteenth century with the development of the mass press. In Paris, in 1836, daily newspapers \textit{Le Siècle} and \textit{La Presse} managed to halve their price relative to competitors by complementing their revenues with the sales of advertising inserts~\cite{gautier2024marchandisation}. 

\subsection{Different Types of Attention}

When it comes to defining attention, several categories should be distinguished.
\textbf{Individual attention} is the ``active direction of the mind upon some object or topic”~\cite{legarrec2019price}. Deciding where to focus one’s attention is not always active and conscious as environmental cues can reflexively attract attention (e.g., a noise in the street or a notification that pops up on one's phone). 
Attention, the process of defining ``useful objects” according to~\citet{bergson2018rire}, is about \textit{focusing} (on something) just as much as it is about \textit{ignoring} (the rest of the environment).

Still, attention is not only an individual process, as it also has communal dimensions that~\citet{citton2016attention} splits into two categories. First, \textbf{joint attention} arises when several individuals interact while being "mutually aware of one another within the same spatiotemporal situation"~\cite{citton2016attention,valette2023trance}. In that setting, their respective attentions are influenced by how they perceive each others' attentional behaviors. For example, if two persons are chatting and one keeps staring at a particular point, it is likely that the other will eventually check whether something worthy of her attention is happening there. Similarly, during a talk in front of an audience, both the speaker and the members of the audience adapt their behavior according to how they interpret cues (e.g., gaze or gestures) of the other party~\cite{citton2018ecologie}. Joint attention is closely related with the sharing of experiences with others~\cite{white2011best}.

Second, \textbf{collective attention} is defined as the attention paid by the members of a community to "the various elements and issues in their environment"~\cite{citton2016attention}. Information spread by newspapers or television deemed important by a large number of individuals typically falls under this category. Collective attention allows humans to act together by focusing on the same object simultaneously~\cite{legarrec2019price}, specifically knowing that others are focused on the same thing at the same moment~\cite{citton2014pour}. Most often, collective attention relies on an infrastructure (e.g., a platform, either digital or physical, such as a newspaper) that concentrates the individual attention of many. Digital tools now enable the unprecedentedly precise measurement of the quantity of attention put on a given focal point (e.g., number of views of a video).
Reified, collective attention no longer belongs to the individuals that produce its unit components. Once aggregated by digital infrastructures and apps, collective attention is monetized by companies. Suffice to say, advertising generates over 95\% of Meta revenues~\cite{saul2023meta}. Thus, firms that rely on selling attention have a clear incentive to collect as much of it as possible.

\bgroup
\def\arraystretch{0.6}
\begin{table}
	\centering
\label{tab:attention_types}
\begin{tabularx}{0.46\textwidth}{L}\toprule
\textbf{Individual Attention}  \\ {\rule[0.01cm]{220pt}{0.5pt}} \\ 
\textit{actively focusing one's mind upon something}  \\ {\rule[0.01cm]{220pt}{0.5pt}} \\ 
e.g., reading a book, looking at a painting, listening to a podcast, writing a paper, etc. \\\midrule

\textbf{Joint Attention} \\ {\rule[0.01cm]{220pt}{0.5pt}} \\ 
\textit{sharing an experience while aware of one another} \\ {\rule[0.01cm]{220pt}{0.5pt}} \\ 
e.g., eating at a restaurant with friends, watching a football game in a stadium, attending a play, etc. \\\midrule 

\textbf{Collective Attention}  \\ {\rule[0.01cm]{220pt}{0.5pt}} \\ 
\textit{knowingly and jointly focusing on the same object}  \\ {\rule[0.01cm]{220pt}{0.5pt}} \\ 
e.g., watching the evening news, watching popular YouTube videos, etc. \\\bottomrule
\end{tabularx}
\caption{Three types of attention}
\end{table}
\egroup

\subsection{Attention as a Commons}

From an economics perspective, individual attention possesses characteristics of goods: it is scarce and tradable~\cite{newman2019regulating}. Accounting for its nature, its importance and its commodification, we argue that \textbf{attention should be defined as a commons}. This idea has been seldom developed in the literature so far. The contribution of~\citet{crawford2015world} who introduced the concept of "attentional commons" is particularly noteworthy. He suggests that framing attention as a "valuable resource that we hold in common", like air or water, will help "figure out how to protect it".~\citet{wagner2015tragedy} identifies that the loss of productivity of knowledge workers due to attentional threats could induce a “tragedy of the attentional commons”. 

In economics, commons (in the sense of common goods) are \textit{rivalrous} (consumption by one agent prevents consumption by others) but \textit{non-excludable} (consumption cannot be prevented).
Commons garnered a lot of attention in the wake of~\citet{hardin1968tragedy} who warned against the "tragedy of the commons" that may arise when agents consume the resource considering their sheer interest, ultimately depleting it~\cite{spiliakos2019commons}. If the rival nature of individual attention is clear, because it can only be consumed by one focus point at a time, the matter of non-excludability is more subtle. Individual attention is theoretically excludable provided that people have their free will and are not manipulated: someone can choose to focus on a book rather than another one. Yet, because attention can be caught against the individual's will, it is not always the case: not paying attention to an individual shouting next to us is an arduous, if not impossible, task. Similarly, in the realm of digital apps, not paying attention to notifications and other attention-catching design tricks is difficult.
At the aggregate level, in the context of digital platforms where content is freely generated by users (theoretically within the limits set by the law and the company), we argue that collective attention is a non-excludable resource unless the platform from which it emerges has designed a mechanism to enable excludability (e.g., by banning or downranking a content creator, such as a newspaper, whose content is deemed not sufficiently engaging).

Maximizing the freedom of individuals to use their attention as they please (putting aside the portion of attention unwillingly attracted by signals of the environment like noises or movement) is a prerequisite for \textit{positive liberty}--the "possibility of acting in such a way as to take control of one’s life and realize one’s fundamental purposes”~\cite{carter2021liberty}. 
In an "online manifesto",~\citet{floridi2015onlife} contends that attention is crucial because “attentional capability is an inherent element of the relational self for the role it plays in the development of language, empathy, and collaboration”.~\citet{watzl2022ethics} stresses that psychology and neuroscience research shows that attention is vital for decision-making, agency, perception, memory, self-control, emotions, and consciousness. Furthermore, he argues that attention is not only important for describing and understanding the mind but should also be a central concern in ethics because a lot can be learnt about someone (character, values, interests, etc.) just by knowing what one pays attention to, how much, and when. Therefore, just like philosophy already studies the ethics of belief, desire, or emotion, \citet{watzl2022ethics} contends that attention deserves similar scrutiny and that, even if it is not always consciously directed, norms can still apply, as they do for emotions. The author proposes several ways to think about the ethics of attention, introducing categories of norms based on the content we pay attention to, the way we pay attention, whether it is instrumental or not to pay attention, and why we care about attention.

All in all, attention must therefore be protected against “hypnotic abdication of reason and will”~\cite{williams2018stand}. In that light, the perspective of the Italian Rodotà Commission on commons excellently characterizes attention. The Rodotà Commission defines commons as things "that express functional utility for the exercise of fundamental rights and the free development of the individual”. Their “collective enjoyment” must be preserved by the law “for the benefit of future generations”~\cite{belenet2020bienscommuns}. 

In this article, we make the point that attention should be conceptualized as a common good and managed as such, that is, including shared, agreed-upon regulation rules.
And we consider that, when it is being given away to an unregulated market, the commodification of attention raises concerns that precisely run counter this common good conceptualization. 
In section~\ref{sec:negative_externalities} we provide an overview of the spectrum of negative externalities of ``attention capture'' and the time spent on screens which justify intervention. In section~\ref{sec:mitigating_externalities} we consider a spectrum of mitigating measures and in section~\ref{sec:tax} we focus on a innovative, promising one: a Pigouvian tax on attention capture, before concluding.

\section{Negative Externalities of Screen Time: Justifications for Public Intervention}
\label{sec:negative_externalities}

Commons are closely tied to the concept of externalities, the “positive or negative outcome[s] of a given economic activity" that affect a third party "not directly related to that activity”~\cite{iisd2025externality}. Environmental pollution is a typical example, as it is often not considered by producers who focus on their private cost rather than the total societal cost incurred by their activities. Negative externalities represent a market failure that justifies public intervention~\cite{helbling2010externalities}.

Social media platforms, on which users aged 16 to 64 worldwide spend over one-third of their time online~\cite{kemp2024socialmedia}, cause growing concern in the public debate--notably regarding their effects on children. Smartphone-Free Childhood, a UK organization campaigning for phone-free childhood, gathered over 100,000 parents in a matter of weeks in 2024~\cite{moshakis2024phonefree}. While almost 90\% of French citizens support a screen ban in nursery schools~\cite{mildeca2024barometre}, an expert commission appointed by French President Macron (the “screens commission”) found “a clear scientific consensus (...) on the harmful effects of screens on several aspects of the somatic health of children and adolescents”~\cite{bousquet2024enfants}.

Adults are not immune to the attention-capturing techniques engineered by platforms nor to the broad range of their harmful effects (i.e., the negative externalities). In France, “between a quarter and a third" of users view their screen time as excessive, and half encounter difficulties for reducing or stopping their digital activities~\cite{mildeca2024barometre}. Beyond individuals, societies as a whole are affected. 

Due to their large number of users, the strength of their network effects~\cite{yoo2025network}, and the great share of the time spent online they account for~\cite{kemp2024socialmedia}, social media platforms are the main focus of concerns regarding the harm caused by attention-capturing techniques. 
Yet, it must be noted that thousands of free apps, popular or not, provide sometimes very modest or futile services for the sheer sake of exposing users to advertisement. In doing so, they participate in a harmful ecosystem that goes beyond social media platforms alone.

Non-exhaustively, harms to the agentivity of individuals, their health, and societal and political harms are key categories of negative externalities caused by the digital attentional markets. 

\textbf{Agency-related harms.} Agency refers to a “person’s autonomous control over his or her actions”~\cite{sokol2015development}. Because conscious awareness and therefore individual identity require the ability to pay attention~\cite{hartford2022attentional}, assaults on attentional resources threaten individuals’ autonomy and life experience~\cite{watzl2022ethics}. This risk is exacerbated by the fact that short-term gratification activities can impact attention~\cite{santos2022association} and constitute behavioral addictions~\cite{berthon2019addictive}. Through tools such as notifications and "likes", social media companies aim to elicit users' compulsive reactions as opposed to conscious and deliberate decisions.
The effect of such tools is even more alarming when it comes to children, for whom the brain's "ability to make a reasoned choice that does not succumb to immediate temptation is not yet fully formed.
When there are too many requests during childhood, a certain decision fatigue sets in, and the subject gives up the fight against the immediate pleasure that the response to an electronic stimulus creates"~\cite{patino2019}.

\textbf{Health-related harms.} A large range of studies highlight various potential harms linked to digital devices and practices. Increased screen time has been associated with heightened ADHD\footnote{Attention-Deficit/Hyperactivity Disorder.} symptoms~\cite{wallace2023screen}. Because more time spent on screens  is correlated to sedentarity, physical health can be affected with, inter alia, links to poor sleep and risk factors for cardiovascular diseases such as high blood pressure or obesity~\cite{lissak2018adverse}. Mental health can also be negatively impacted. A meta-analysis finds that around one in four children and young people display “problematic smartphone usage” with “at least some element of dysfunctional use, such as anxiety when the phone was not available or neglect of other activities"~\cite{sohn2019prevalence}. A relationship between problematic smartphone usage and deleterious mental health symptoms (e.g., depression, anxiety, high levels of perceived stress, poor sleep) is identified.

\textbf{Political harms.} Social media platforms have often been recognized as causes of increased political polarization~\cite{barrett2021techpolarization} and threats to democracies through “foreign influence campaigns, intentional dissemination of misinformation, and incitements to violence”~\cite{omidyar20186}. Scholars, reporters, and United Nations investigators alike have found that Facebook contributed to the 2017 genocide of Rohingyas in Myanmar, hosting a large amount of misinformation and hateful content targeting the Rohingyas~\cite{zaleznik2021facebook}. 
As another example, with the increasing popularity of video games, gaming and adjacent platforms have been increasingly instrumentalized to spread extremist propaganda~\cite{lakhani2021video,europol2021european,schlegel2021extremists}.

\textbf{Societal harms.} The attention economy is linked to, and risks worsening, social inequalities. In places where most people can access the internet, socioeconomic vulnerability can be linked to increased time on social media or digital games~\cite{hartford2022attentional}. Money is more and more frequently required to reclaim back one’s attention, which risks furthering attentional inequalities, leading~\citet{crawford2015cost} to argue that silence has turned into a luxury good. Research suggests that a less advantaged socioeconomic context is a risk factor for developing internet addiction~\cite{hartford2022attentional}. Considering educational attainment as an indicator of socioeconomic status,~\citet{scheerder2019negative} find that, while exposure to negative aspects of the internet is similar for all users, members of highly educated groups try to take control through remedial actions, whereas their less educated counterparts tend to experience negative outcomes more passively.


All these harmful consequences of digital devices and experiences constitute negative externalities in the economic sense. The link is explicitly made by some authors, yet quite rarely to the best of our knowledge:~\citet{puig2023societal} argues that “online polarization is a negative externality of algorithmically mediated platforms”.~\citet{carnovale2022online} explain that “online platforms are generating advertisement space whose market is producing a negative social externality in the form of a negative mental health impact on users”.~\citet{verveer2019countering} refers to the damages to democracies caused by “foreign influence campaigns, intentional dissemination of misinformation, and incitements to violence inadvertently enabled by (...) digital platform companies” as another negative externality.

Besides, despite a growing consensus among academics~\cite{europeanparliament2023addictive}, whether behavioral addictions, such as compulsive use of social media, are ”true" addictions is still debated~\cite{berthon2019addictive}. 
Still, the negative externalities highlighted in this section call for immediate regulation of the attention economy~\cite{helbling2010externalities}. 
In the European Union, the lack of regulation related to addictive design was recently underscored by the~\citet{europeanparliament2023addictive} which identifies addictive design features (e.g., "pull-to-refresh" page reload) and reasons why certain mechanisms can lead to addiction (e.g., intermittent variable rewards inducing dopamine surges). 
Providing an overview of existing legislation,~\citet{albrecht2023digital} also contend that the problem of digital addiction caused by persuasive design is insufficiently addressed in the European Union. 
Arguing against the idea that reforming the digital attention economy would be premature because of a lack of certainty about "the precise causal relationships between particular designs and particular types of harm",~\citet{williams2018stand} explains that the nature of attentional markets (e.g., the pace of technology evolution, the diverse relationships of users to technology, etc.) renders the establishment of definite knowledge about their effects difficult (when compared to drug consumption for example).
The author likens the context to requiring verification that “the opposing army marching toward you do, indeed, have bullets in their guns”. In other words, even through a very cautious perspective, regulation is justified on the grounds of the precautionary principle, which presupposes that "potentially dangerous effects deriving from a phenomenon, product or process have been identified, and that scientific evaluation does not allow the risk to be determined with sufficient certainty"~\cite{EuropeanCommission2000_PrecautionaryPrinciple}.
Considering the "urgency" of the situation and the risk that massively and globally wasting attention may not be "sustainable" for a civilization,~\citet{michel2024pay} propose a set of principles to steer the design of applications in order to regulate attention capture.


\section{Mitigating Externalities:\\ The Different Levers for an Intervention}
\label{sec:mitigating_externalities}

Since the digital harvesting of human attention causes negative externalities, which justify public intervention~\cite{helbling2010externalities}, the next question to consider is the nature of the mitigation response that governments should implement.~\citet{berthon2019addictive} identify three categories of policy solutions to protect consumers against digital addiction: \textit{Inform}, \textit{Restrict} and \textit{Guide}. These options are illustrated in Table~\ref{tab:inform-guide-restrict} with the example of infinite scrolling provided by~\citet{albrecht2023digital}. It is important to mention that the categories are not mutually exclusive as there can be some overlap~\cite{albrecht2023digital}.
%
%


In the rest of this section, we describe and illustrate further the three types of policies.

\bgroup
\def\arraystretch{0.6}
\begin{table}
	\centering
\begin{tabularx}{0.46\textwidth}{L}\toprule
\textbf{Inform}  \\ {\rule[0.01cm]{220pt}{0.5pt}} \\ 
\textit{enhancing transparency and the users' knowledge}  \\ {\rule[0.01cm]{220pt}{0.5pt}} \\ 
e.g., mandating platforms to provide users with information about the effects of endless scrolling \\\midrule

\textbf{Restrict}  \\ {\rule[0.01cm]{220pt}{0.5pt}} \\ 
\textit{banning particular features viewed as too harmful or limiting digital devices use by vulnerable individuals}  \\ {\rule[0.01cm]{220pt}{0.5pt}} \\ 
e.g., banning the infinite scrolling feature \\\midrule

\textbf{Guide} \\ {\rule[0.01cm]{220pt}{0.5pt}} \\ 
\textit{encouraging or incentivizing a change of behavior}  \\ {\rule[0.01cm]{220pt}{0.5pt}} \\ 
e.g., granting users the right to deactivate or activate the infinite scrolling feature at any time, with deactivation set as the default option \\\bottomrule 
\end{tabularx}
\caption{Three categories of policy solutions to protect consumers against digital addiction: \textit{Inform}, \textit{Restrict} and \textit{Guide}}
\label{tab:inform-guide-restrict}
\end{table}
\egroup

\subsection{\textit{Inform} Measures}

\textit{Inform} measures aim to enhance the transparency and the knowledge of users. This may take the form of requiring platforms to provide information to users (e.g., about the effect of endless scrolling on time spent on the platform, about the dangers of using devices that emit blue light during nighttime).
More generally, for goods built to trigger compulsive use, \textit{Inform} measures could be inspired by some principles proposed by~\citet{michel2024pay}. The \textit{principle of continuous reflexivity} suggests that "users must be provided a continuously updated feedback on their usage of the system and on themselves to support their reflexivity and maintain an up-to-date informed consent". Additionally, the \textit{principle of full user awareness} proposes that "users must be made aware of all the features and purposes leading to a recommendation, before and when it is provided".

The effect of information-related policies on individual behavior can be limited as highlighted by the case of smoking~\cite{shadel2019graphic,strong2021effect}. 
However, while \textit{Inform} measures alone can be insufficient, they still constitute an important complement to other strategies.
For instance, graphic warning labels may enhance the efficiency of other control strategies to reduce cigarette smoking; an informative measure may prepare the citizens for debating a regulatory measure~\cite{strong2021effect}. 

\textit{Inform} measures could also be designed to address the broader scope of issues surrounding artificial intelligence (AI), which encompasses, but is not limited to, attention capture. Indeed, the attention economy now heavily relies on AI algorithms, such as those used in recommendation systems by social media platforms to deliver content that more effectively capture users' attention~\cite{haque2025explain}.
In this regard, we propose the idea of an AI-Score, a Nutri-Score-inspired system\footnote{Nutri-Score is a "front-of-pack label that provides user-friendly information on the nutritional quality of food and beverages"~\cite{iarc2021nutriscore}.} for assessing AI applications' respectfulness of their users.
The score would be designed as a simple, color-coded, and letter-based front-of-app label that could typically be displayed in application stores. The aim is to help users quickly assess how respectful and safe an AI application is regarding user autonomy, privacy, and well-being. This would raise public awareness, bolstering up open debate and the crafting of complementary regulation strategies. 
\begin{figure}[h]
    \centering
    \includegraphics[width=0.472\textwidth]{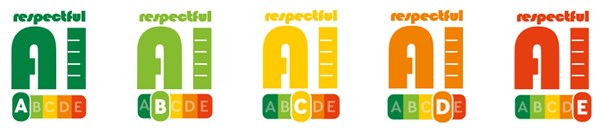}
    \caption{AI-Score: a Nutri-Score-inspired system for assessing AI applications' respectfulness of their users}
    \label{fig:AIScore}
\end{figure}

The score should capture both positive and negative aspects of an AI application's design and usage, such as the ones identified in~\citet{michel2024pay}. 
Positive factors would increase the score and include desirable properties such as explainability, traceability, and auditability.
They would also include features such as opt-in by default, session time limits, the availability of focus or "do not disturb" modes, clear consent and process flows, etc.
On the other hand, negative factors would decrease the score and account for features such as dark patterns and attention traps (e.g., autoplay, infinite scrolling, frequent or unsolicited notifications), emotional manipulation (e.g., content optimized to trigger negative emotions such as anger or indignation), forced opt-ins, opaque algorithms, hidden tracking, etc. The balance of these elements would produce an overall score falling into one of the following five categories. 

\noindent ``\textbf{\color{OliveGreen}A}'' would correspond to an exemplary design that is time-respecting, emotion-neutral, and fully transparent.

\noindent ``\textbf{\color{Green}B}'' would indicate a responsible design with minor attention hooks and clear user controls.

\noindent ``\textbf{\color{Dandelion}C}'' would call for caution in applications that include some addictive design patterns but provide basic safeguards.

\noindent ``\textbf{\color{Orange}D}'' would indicate a problematic design featuring dark patterns and limiting user autonomy.

\noindent ``\textbf{\color{Red}E}'' would stamp exploitative applications that are emotionally manipulative, data-opaque, and pose high addiction risks.

\noindent 
In turn, such an \textit{Inform} measure could incentivize developers to adopt more ethical practices when designing AI applications in order to attract users, similarly to how the Nutri-Score may encourage food reformulation~\cite{steenbergen2024comparison}.

\subsection{\textit{Restrict} Measures}

\textbf{\textit{Restrict}} measures seek to ban particular features viewed as too harmful or to prevent or limit the use of digital devices by vulnerable individuals, such as children, in order to protect them. They could take the form of laws forbidding the use of certain business models or specific design patterns by platforms. Policies that limit the maximum amount of daily time one can spend on online platforms would also fall into this category.
Many restrictive measures have been introduced over the last decade with the aim of protecting young people.
In 2015, Taiwan made it illegal for children under the age of two years old to use screens~\cite{duncan2015ipads}. Australia recently amended the Online Safety Act to ban the usage of social media for children below 16 years~\cite{FARDOULY2025e235}.
According to~\citet{unesco2023}, "almost one in four countries has introduced such bans \textit{[of mobile phones in schools]} in laws or policies". Some countries, such as Japan, Belgium, or China, have introduced legislation against loot boxes in video games because of how similar they are to gambling~\cite{gonzalez2023unregulated}. 

While there is consensus when it comes to protecting vulnerable groups such as children, implementing similar measures targeting the average population may be more challenging in Western societies that strongly value individual freedom. For example, according to a recent survey requested by the Joint Research Center of the European Commission, self-direction is one of the two most important personal values for the citizens of the European Union~\cite{becuwe2021special}. 
Besides, companies tend to oppose ideas revolving around individual freedom and consumer responsibility as a strategy to undermine regulation attempts~\cite{korn2003framing,berthon2019addictive}.
In the United States, such behaviors are often framed as "a moral weakness and a matter of willpower"~\cite{bernhard2007voices,berthon2019addictive}.

\subsection{\textit{Guide} Measures}

\textbf{\textit{Guide}} measures aim to encourage or incentivize a change of behavior. Contrary to \textit{Restrict} measures, they are not about outright prohibition. This may render them especially appropriate for protecting fundamental rights and political freedoms, which is a crucial element for regulators to account for~\cite{arcila2023social}. A first category of \textit{Guide} measures involves changes to the platform's interface aimed at modifying the user experience. For instance, this could mean giving users the ability to activate or deactivate a feature such as autoplay at any time.
This could also mean deactivating recommendations or attention-fragmenting features (e.g., notifications) by default.
\citet{michel2024pay} generalize this type of measure with the \textit{principle of due diligence}, proposing that users "should always be given and made aware of the options to escape the systems’ loops, processes and goals".

Another category of \textit{Guide} measures consists of incentives for companies, typically in the form of fiscal measures.
In this respect, \citet{newman2019regulating} proposes to limit the amount of advertising expenditures that can be deducted from companies' revenues to alleviate their taxes.
This is in line with the \textit{principle of the right incentive} put forth in~\citet{michel2024pay}, that states that "governance bodies should leverage legal and economic means to drive platforms’ practices towards desirable behaviours, while penalizing undesired behaviours".

All in all, based on this review,
it seems essential to leverage all three types of measures.
Yet, we argue that \textbf{the strengths of \textit{Guide} measures make them the core components of the action plan that policymakers should undertake,} while \textit{Inform} and \textit{Restrict} measures should complement and support them, as exemplified above.
This is aligned with~\citet{albrecht2023digital} who identify the \textit{Guide} option as the "most favourable" in their evaluation of the three options against three criteria---enabling digital self-determination, regulatory feasibility, and room for business and innovation---in the context of digital addiction in the European Union.

\section{Taxing to Reduce Externalities: for a Pigouvian Tax on Attention Capture}
\label{sec:tax}

Within the category of \textit{Guide} measures considered by~\citet{berthon2019addictive}, \textbf{taxation is particularly promising}.~\citet{puig2023societal} argues that taxation is a better alternative than other policies such as content moderation or algorithm changes. Indeed, despite laws that may require platforms to enforce appropriate content moderation,  it is likely that companies never make the necessary investments as those would be opposed to their financial interest, for instance, to spread divisive (thus engaging) content~\cite{michel2024pay}.
By contrast, a tax would act as a “financial disincentive” to deter the building of “algorithms whose feedback loops and unintended consequences result in polarization”~\cite{puig2023societal}. 
The relevance of a tax is reinforced by~\citet{newman2019regulating} who argues against the creation of “attention rights” that would lead to an intractable conflict with speech rights (e.g., an individual seeking to assert her attention rights against a company wishing to assert its commercial-speech rights by advertising).
Instead, \citet{newman2019regulating} suggests to rethink existing antitrust and fair competition laws, and discusses taxing corporate attention consumption as this would ``disincentivize attention intermediaries from vacuuming up as much attention as possible'', and is ``justified on general tax-law grounds or, alternatively, as a Pigouvian measure''.

\subsection{Making the Case for a Pigouvian Tax}

In economics, taxation is a typical answer to negative externalities. Named after British economist Arthur Pigou~\cite{sandmo2016pigouvian}, Pigouvian taxes are designed to “internalize externalities” and reduce the volume of consumption of a given good to the \textit{socially optimal level}. The advantages of Pigouvian taxes have been highlighted by many researchers~\cite{masur2015toward}. Founded by the leading economist Gregory Mankiw, the Pigou Club, now encompassing many prominent scholars~\cite{mankiw2006rogoff}, advocates in favor of Pigouvian taxes because of their benefits~\cite{mankiw2009smart}.

In line with these works, we argue that implementing a Pigouvian tax on digital platforms is a promising option for regulating attention markets. 
First, when a polluter pays for the pollution it engenders, it is compelled to internalize the cost of pollution by incorporating it into its cost function. 
Hence, a Pigouvian tax on digital platforms would make them account for the negative externalities of their activities while setting a price on the attention and time of users. This would effectively remove companies’ incentive to harvest as much attention as possible, and may foster a change in their behavior. For example,~\citet{acemoglu2024urgent} argue for a tax on digital advertising aimed at encouraging companies to shift their business models towards alternatives not "based on keeping people addicted and sustaining intense emotional responses", such as subscription-based models.
Second, setting up a tax raises revenue that can be used to lessen the harms caused. For instance, taxing sugar-sweetened beverages can fuel investments in the healthcare system or compensate for the social burden, notably by funding health- and education-related programs~\cite{bayliss2018sweet}.
Likewise, the revenue collected from a Pivouvian tax on attention-capturing services could be used to invest in educational projects meant to raise public awareness, or compensate for the cost of worsened mental health. 
Third, while attention is currently harvested for “free”, considering a Pigouvian tax would provide an opportunity to initiate a public debate and spur awareness about the free, unfettered access to individual attention and time.
Leveraging the concept of attentional commons discussed in section~\ref{sec:nature_of_attention}, the debate could emphasize the value of attention and time, putting forth the idea that these resources do not need to be free and that regulation can be implemented.

\subsection{Applying a Pigouvian Tax on Attention Capture}

\begin{figure}[h]
    \centering
    \includegraphics[width=0.47\textwidth]{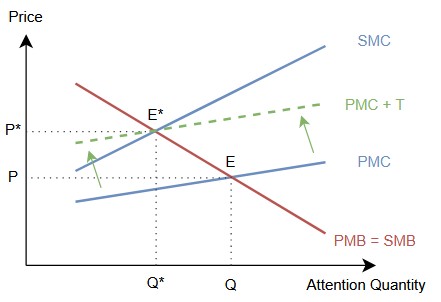}
    \caption{Supply and demand diagram of a Pigouvian tax on attention markets. 
    On the supply side, the private marginal cost ($PMC$) is the cost incurred by platforms to harvest one additional unit of attention. 
    On the demand side, the private marginal benefit ($PMB$) is the utility that advertisers obtain from one additional unit of attention.
    The social marginal cost ($SMC$) is the cost, for society, of harvesting one additional unit of attention, while the social marginal benefit ($SMB$) is the benefit, for society, generated by the consumption of one additional unit of attention. Without regulation, the equilibrium $E$ represents the quantity $Q$ of attention exchanged at price $P$. After introducing the Pigouvian tax $T$, the equilibrium $E^{*}$ is reached with a quantity $Q^{*}$ of attention exchanged at price $P^{*}$. The private marginal cost with the Pigouvian tax imposed on platforms is $PMC + T$.} 
    \label{fig:attention_market}
\end{figure}

As a first approach, we propose to visualize the effect of a Pigouvian tax on attention markets with a supply and demand diagram (Figure~\ref{fig:attention_market}), common in economics.
In this simplified setting, digital platforms are the suppliers of a resource: the aggregated attention obtained from individual users. 
Demand emerges from advertisers who covet this attention.
The total amount of attention available for exchange is finite; advertisers must compete to obtain it (due to attention being a rivalrous good) with mechanisms such as real-time bidding.

The private marginal cost ($PMC$) is the cost incurred by platforms to harvest one additional unit of attention\footnote{Digital platforms are often described as having "near-zero marginal costs"~\cite{aurecon2025zero} because providing the service to additional users costs almost nothing. However, the marginal cost of capturing additional units of user attention is likely to eventually rise as the market becomes saturated, because attention is scarce and rivalrous. Consequently, we represent $PMC$ as an increasing function.}.
The social marginal cost ($SMC$) represents the additional cost for society entailed by harvesting one more unit of attention. As a first approximation, we assume that it encompasses the negative externalities described in section~\ref{sec:negative_externalities}.

The equilibrium of an economic market is the point where supply meets demand.
It is reached when the private marginal benefit ($PMB$) equals the private marginal cost ($PMC$).
On Figure~\ref{fig:attention_market}, the equilibrium is reached at the point $E$ characterized by a quantity $Q$ and a price $P$.
Let us briefly explain how the market converges to this equilibrium point. On the left of point $E$, $PMB>PMC$: the marginal benefit for advertisers of consuming an additional unit of attention---i.e., their willingness to pay---is greater than the cost of producing it for the platform. As a consequence, the platform profits from selling more attention. On the right of point $E$, $PMB<PMC$, the sale of an additional unit of attention would result in a loss for the platform because the marginal benefit is inferior to the marginal cost.

Without regulation, the private marginal cost ($PMC$) incurred by platforms is inferior to the social marginal cost ($SMC$) generated by their activity.
As a consequence, the equilibrium quantity $Q$ of attention consumed is above the level that would be socially optimal, because it is sold at a price $P$ below the socially optimal one.
Public authorities react by introducing a Pigouvian tax imposed on platforms. The tax $T$ should be calibrated so that it equals the difference between the social marginal cost and the private marginal cost for the socially efficient quantity $Q^{*}$~\cite{mcafee2012beginning}, i.e., 
\[
T = SMC_{Q^{*}} - PMC_{Q{*}} 
\]

\noindent The tax $T$ imposed on platforms pushes the supply curve ($PMC$) upward, resulting in a new supply curve ($PMC + T$). At the new equilibrium $E^{*}$, the quantity of attention consumed is reduced to the socially efficient level $Q^{*}$, and it is exchanged at a higher price $P^{*}$.

To build on this simple model, further research is required to better account for the specificities of the attention markets, and to depict more precisely the effects of a Pigouvian tax.
For instance, analyzing how the burden of the tax would be split between the two sides of the market (i.e., platforms and advertisers) is important. Indeed, the statutory incidence of a tax differs from its economic incidence~\cite{fullerton2002tax}. The former relates to the legal responsibility of “writing the check"~\cite{fox2022statutory} while the latter is about who actually bears its cost. The distribution of the tax burden between digital platforms and advertisers will depend on price elasticities, namely how supply or demand evolves as a reaction to price change.
Furthermore, the design of the tax itself raises challenges, most notably the estimation of its amount and the definition of its scope. In the next subsections, we discuss these challenges and outline the research leads, that, we believe, should be investigated.

\subsection{Measuring Harm to Calibrate a Pigouvian Tax}
\label{sec:measuring}

A first challenge in designing the proposed Pigouvian tax lies in measuring, qualitatively and quantitatively, the individual and collective harms caused by the capture of attention by platforms. 
The qualification of harms should consider the categories described in  section~\ref{sec:negative_externalities}, and should also leverage and extend existing works such as the taxonomy of harms caused by dark patterns, proposed by~\citet{santos2025no}. 
On the other hand, quantifying harms is a difficult enterprise, notably when they affect the psyche of individuals~\cite{baumol1972taxation}. 
In the case of attention markets, ascribing a monetary value to time is another conundrum~\cite{kondylis2022time}. In addition, a theoretically exact approach would not only imply measuring negative externalities but also positive ones~\cite{salib2021pigouvian}, such as opportunities to access information. In this respect, \citet{arcila2023social} identify the challenge of regulating social media platforms to mitigate threats while preserving their benefits.
In any case, it could be beneficial to build upon previous works examining the quantification of externalities in different contexts, such as cybercrime~\cite{khan2015every}, the presence of wind turbines~\cite{krekel2017does}, or outdoor advertising~\cite{czajkowski2022valuing}. 

Without an exact quantification of harms, it would still be possible to use approximations or proxies. In the case of climate change mitigation, many countries have introduced carbon taxes as a remedy for the negative externalities caused by carbon emissions~\cite{brzezinski2025carbon}, despite the difficulty of evaluating damage~\cite{pezzey2019social}.
Arguing that an incentive-based approach may be more effective than a command-and-control policy,~\citet{carnovale2022online} propose a mechanism to prevent applications from enticing users into spending an “excessive amount of time” and protect their mental health. For all users in a given jurisdiction, for whom time spent is greater than an agreed-upon threshold, the platform would have to pay an amount that increases with the time spent, as per the following, \[
T = \sum_{i=1}^{N} t(x_i) \quad \text{\textnormal{where}\ } 
t(x_i) = \left\{
\begin{array}{ll}
0, & x_i < \bar{x} \\
f(x_i - \bar{x}), & x_i \geq \bar{x}
\end{array}
\right.
\]
where $T$ is the amount of the tax, $N$ the number of users, $x_i$ the time spent by user $i$ and $\bar{x}$ the threshold.~\citet{carnovale2022online} note that the tax may incentivize companies to change their behavior. For instance, they might stop sending notifications to users if they are close to the time limit $\bar{x}$. 
Considering the difficulty of ascribing a monetary value to the potential harm on mental health induced by the excessive use of platforms, the authors propose a tax proportional to the revenues generated by an additional time unit of average user engagement (captured in the $f$ function) rather than a purely Pigouvian tax that would take into account the actual social cost of the externality.

Beyond this proposal, various other parameters could be used to estimate harm severity as alternatives to exact quantification. In addition to directly measuring polarization footprints,~\citet{puig2023societal} suggests several proxies for taxing online polarization, such as taxing data centers or data brockerage. 
Many other proxies can be identified, and they could also be combined. 
For instance, it is well established that the notifications raised daily by multiple applications fragment attention, thus hampering focus and causing error-prone contexts. Similar to the proposition by~\citet{carnovale2022online}, a tax could be calibrated to increase with the number of notifications sent to users beyond a given threshold, thus leading to an extension of the formula of Carnovale and Ramirez:
\[
T = \sum_{i=1}^{N} t(x_i)+t_n(n_i) \quad  \text{\textnormal{where }\ }  \]

\[
t_n(n_i) = \left\{
\begin{array}{ll}
0, & n_i < \bar{n} \\
f_n(n_i - \bar{n}), & n_i \geq \bar{n}
\end{array}
\right.
\]
where $n_i$ is the number of notifications sent to user $i$ and $\bar{n}$ the threshold.

\subsection{Defining the Scope of a Pigouvian Tax}
\label{sec:scope}

A second challenge lies in defining the scope of the services that shall be concerned by the tax. 
One approach consists in simply working around this challenge by making recommendations with respect to vulnerable populations.
Previous studies have quantified recommended screen time for children as a function of their age. For example, the French screens commission advises not to expose children under 3 years of age to screens~\cite{bousquet2024enfants}. The~\citet{aacap2024children} limits “non-educational screen time to about 1 hour per weekday and 3 hours on the weekend days” for children between 2 and 5. 

There is however less guidance for older individuals.
It has been suggested that individuals “should spend at least three to four hours each day completely detached from screens”~\cite{serrano2022experts}. The American~\citet{nhlbi2013reduce} advises to limit screen time at home to two hours or less a day, except if it is work- or homework-related. 
However, going beyond such general recommendations requires to take into consideration that not all screen time is worth the same harm or benefit, and breaking down time raises difficulties.
Typically, a quality course on YouTube may be deemed more enriching than the same time spent browsing entertaining short videos on social media. Yet, the point is not to oppose educational versus entertaining content, as both have their benefits, nor to oppose "essential" versus "non-essential" internet uses~\cite{berthon2019addictive}.
Rather, we should focus on how to break down screen time, since experts contend that time spent on social media apps is the most concerning~\cite{serrano2022experts}, and that excessive online social media use is more strongly associated with self-harm behaviors, depressive symptoms, low life satisfaction, and low self-esteem than electronic gaming or television watching~\cite{twenge2021not}.
Conversely, 
a study suggests that limiting Facebook, Instagram, and Snapchat use to 10 minutes per platform and per day for three weeks is associated with significant reductions in loneliness and depression~\cite{hunt2018no}.

These elements underline the importance of defining the regulation's scope---that is, which digital products should be subject to the tax. Noting that a large majority of users’ time on social media is spent on a handful of platforms with a large user base (e.g., Facebook, Instagram, YouTube, Snapchat, and X),~\citet{albrecht2023digital} argue that policy responses to digital addiction should target Very Large Online Platforms (VLOPs), defined in the Digital Services Act (DSA) as platforms with more than 45 million monthly users in the European Union.
The authors propose that, for each VLOP, regulators assess its contribution to digital addiction by using selected metrics as proxies (e.g., average daily usage time, daily usage time of the 20\% most active users), with the platform becoming subject to the new regulation if these proxies exceed agreed-upon thresholds.
The DSA already imposes a number of specific obligations on VLOPs due to their size and potential impact on society~\cite{ec2023dsa}. They are required to ``identify, analyse, and assess systemic risks that are linked to their services'' and to ``put measures in place that mitigate these risks'',  for instance by adapting the design of their services or modifying their recommender systems.
Furthermore,~\citet{albrecht2023digital} point out that complying with regulation against digital addiction could imply a substantial cost, which VLOPs are better placed to bear. This point is worth considering to avoid penalizing relatively smaller service providers, as this would bear the risk of further concentrating digital markets in the hands of a few dominant players~\cite{unctad2025globaltrade}.

In the case of the attention economy, such asymmetrical regulation may create a weak point since platforms with a user base below the agreed-upon threshold would avoid taxation even if they apply highly exploitative, manipulative or addictive techniques. Besides, were this approach implemented, companies with a number of users above the threshold might fragment their offerings into several products to dodge the policy, even though there might still be improvement because transitions between apps could represent natural stopping points and decrease overall use~\cite{carnovale2022online}. 

Two key points should be extracted from this analysis.
First, there is a need for multidisciplinary studies to better understand how digital companies may respond to taxation and to specific tax designs. As another example, let us suppose that the tax amount increases with the number of ads shown. Digital platforms might then decide to show more ads to users who click on ads most often, raising ethical concerns. 
Second, accounting for the size of a platform's user base is important to determine which companies should be subject to the tax, particularly because, as a result of network effects, the value of the service increases with the number of users~\cite{yoo2025network}. Yet, this metric should be complemented with a broader set of indicators to capture additional aspects of the platform’s impact and functioning.

Taken altogether, these metrics could be used to classify each digital product whose business model relies on attention harvesting according to an \textbf{Attention Harvesting Scale}.
The scale could take inspiration from the four levels of risk (unacceptable, high, limited, minimal) laid out by the AI Act to classify AI systems~\cite{edwards2021eu,europeancommission2025aiact}. It should account for the harms and risks associated with various practices, while laying the groundwork for calculating the amount of the tax.
%
%
Let us add that, although the scale we propose would apply to digital services, it could be expanded to encompass other advertisement infrastructures, e.g., television channels and outdoor advertising, and associated negative externalities such as over-consumption or visual pollution~\cite{czajkowski2022valuing}.
Meanwhile, through a cross-fertilization approach, regulating the attention economy should take inspiration from the advertising laws that most countries introduced to protect consumers~\cite{michel2024pay}.


\section{Conclusion}
\label{sec:conclusion}

The global and unprecedented harvesting of attention, a pillar of human experience and independence, is of the utmost concern. Affecting both individuals and societies, the broad, negative externalities associated with attention-capturing platforms encompass reduced individual agency, adverse health outcomes, and societal and political harms such as democratic erosion and inequality. Urgent public intervention is now required to protect attention, a common good threatened by over-exploitation.

In this paper, after a review of the policy options that should be leveraged by policymakers to mitigate these issues, we have identified the principle of a Pigouvian tax imposed on attention-harvesting companies as a promising lead to protect individuals and societies. 
Such a tax would incentivize companies to adjust their business models while protecting user autonomy.
We have discussed structuring elements that should help answer the core questions behind such a policy, e.g., how to design the tax, which companies and digital services should be targeted, and the need for an Attention Harvesting Scale. To the best of our knowledge, no country has introduced such a policy so far, and, while Uganda did create a tax on social media in 2018, it targeted users rather than companies and aimed more at repressing critical voices than protecting individuals~\cite{boxell2022taxing}.

Nevertheless, there is a compelling precedent for using taxation as a means to safeguard attentional resources: noise taxes, designed to correct the adverse effects (i.e., the negative externalities) of noise pollution~\cite{ezcurra2018noise}. In France, airlines have to pay this tax in a number of airports~\cite{cidb2019nuisances} and the revenues are used to finance noise pollution mitigation measures. Interestingly, just like attention, silence, the resource that noise taxes aim to protect, has been framed as a commons by~\citet{illich1982silence}, who had already identified that silence, “necessary for the emergence of persons (...) is taken from us by machines that ape people”. Unsurprisingly, noises are one of the most common methods used by social media platforms to attract users’ attention (e.g., the sound linked to a notification).

Let us finish with an analogy. If we visualize the sum of our attentional resources as a lively and beautiful forest, it is about time for lumberjacks to start paying for the trees that they relentlessly collect for free, or the whole ecosystem risks crumbling.

\section*{Acknowledgements}
\label{sec:acknowledgements}
This work was supported by the French government through the France 2030 investment plan (ANR-22-CPJ2-0048-01) and the 3IA Cote d'Azur (ANR-23-IACL-0001).

\bibliography{aaai25}

\end{document}